**Stark Tuning and Electrical Charge State Control of Single Divacancies in Silicon Carbide**

Charles F. de las Casas[1], David J. Christle[1], Jawad Ul Hassan[2], Takeshi Ohshima[3], Nguyen T. Son[2], David D. Awschalom[1]

[1]Institute for Molecular Engineering, University of Chicago, Chicago, Illinois 60637, USA

[2]Department of Physics, Chemistry and Biology, Linköping University, SE-581 83 Linköping, Sweden

[3]National Institutes for Quantum and Radiological Science and Technology, 1233 Watanuki, Takasaki, Gunma 370-1292, Japan

**Abstract**

Neutrally charged divacancies in silicon carbide (SiC) are paramagnetic color centers whose long coherence times and near-telecom operating wavelengths make them promising for scalable quantum communication technologies compatible with existing fiber optic networks. However, local strain inhomogeneity can randomly perturb their optical transition frequencies, which degrades the indistinguishability of photons emitted from separate defects, and hinders their coupling to optical cavities. Here we show that electric fields can be used to tune the optical transition frequencies of single neutral divacancy defects in 4H-SiC over a range of several GHz via the DC Stark effect. The same technique can also control the charge state of the defect on microsecond timescales, which we use to stabilize unstable or non-neutral divacancies into their neutral charge state. Using fluorescence-based charge state detection, we show both 975 nm and 1130 nm excitation can prepare its neutral charge state with near unity efficiency.

Optically active point defects in semiconductors are promising candidates for a variety of technologies, from compact sensors of temperature[1], magnetic[2] and electric fields[3], to nodes in a quantum information network[4]. Although the nitrogen vacancy (NV) center in diamond has historically attracted the most attention, recently other defects in different materials have begun to be investigated for their potentially superior properties.[5–7] Among these, the neutrally charged divacancy ($V_C V_{Si}$) in silicon carbide has emerged as a promising alternative system that resides in a host material that is amenable to a variety of microfabrication techniques and is commercially available in 6-inch wafers.[6] However, unlike diamond NV centers, which emit at visible wavelengths, divacancies in SiC emit near telecom wavelengths[6], making them subject to significantly lower loss in optical fiber (0.7 dB/km vs 8 dB/km) and thus more promising for integration with existing optical fiber networks.

Recently, the divacancy was shown to possess a set of spin-selective optical transitions that can coherently link the divacancy's electronic spin to light and are an essential precursor for creating a quantum network composed of divacancy spins.[8] These transitions connect specific sublevels of the orbital ground and excited states (GS and ES) and can be used to initialize and detect its spin state with high fidelity at temperatures below 20 K.[8] Remote entanglement protocols function by interfering indistinguishable photons emitted when these transitions are excited in separate defects.[9] However, their indistinguishability is degraded by variations in strain that perturb the optical transition frequencies randomly by several GHz. Strain inhomogeneity also frustrates their coupling to optical cavities, which are useful for enhancing their emission into the zero phonon line (ZPL) and increasing the production of indistinguishable photon pairs.[10,11] A method for tuning the divacancy's optical resonances is thus crucial for generating remote entanglement between different defects.

Here we show that the optical transitions of neutral divacancies in 4H-SiC can be tuned electrically via the DC Stark effect, which provides a functional way to counteract solid-state inhomogeneity. In addition, we demonstrate fast electrical control of the divacancy defect's charge state. We further demonstrate that we can electrically stabilize the neutral charge state of divacancies that naturally exist in a dark charge state, and determine that both 975 nm and resonant 1130 nm excitation can be used to prepare the neutral charge state of the divacancy with near unity efficiency when under appropriate electrical bias. This compares favorably to the NV center in diamond, which ionizes under resonant (637 nm) excitation and typically relies on a second (532 nm) laser to periodically reset the defect to its proper negatively charged state with only 75% efficiency.[12]

We perform these experiments on a 120 µm thick single-crystal epitaxial film grown on an n-type 4H-SiC substrate using hot-wall chemical vapor deposition.[13] The sample preparation is described in detail elsewhere.[14] We isolate single divacancy defects in a home built confocal microscope using off-resonant 975 nm excitation and perform photoluminescence excitation spectroscopy using a narrow line resonant laser that can be tuned across the ZPL transitions near 1130 nm while detecting single photons from the sideband fluorescence using a superconducting nanowire detector as shown in Fig 1(a). We then use lithographically patterned Ti/Au electrodes on the sample surface to apply electric fields to the defects while measuring the transition frequencies.

We focus on the (hh) and (kk) forms of divacancy, which are oriented along the c-axis of the 4H-SiC lattice, because the structure of their spin-1 $^3E$ excited states and $^3A_2$ ground states has recently been established[8], and they benefit from a higher symmetry relative to the basal-oriented (kh) and (hk) divacancy forms.[8,15,16] Their $^3E$ excited state has six levels whose precise energies and spin character are determined by the spin-orbit and spin-spin interactions intrinsic to the ES manifold. These levels are further perturbed by both strain and electric fields. The electric field perturbation (Stark effect) to the Hamiltonian is given by $\hat{H}_{Stark} = -\vec{d} \cdot \vec{F}$, where d is the electric dipole moment and F is the electric field. We focus on spin conserving transitions between the GS orbital singlet |$A_2$; $S_z$=0> and two ES orbital eigenstates |$E_x$; $S_z$=0> and |$E_y$; $S_z$=0> which are degenerate in the absence of strain and electric fields.[8,17] These transitions are cycling transitions and thus play an important role in fluorescence-based detection of the ground state spin. Since the ground state can also experience DC Stark shifts, the frequency shift of the transitions is determined by the relative change in electric dipole moments of the ES and GS, $\Delta d_{||} = d_{||ES} - d_{||GS}$ and $\Delta d_\perp = d_{\perp ES} - d_{\perp GS}$. The effect of Stark shifts on these transitions is given by $h\Delta v_{E_x} = \Delta d_{||}F_{||} + \Delta d_\perp F_\perp$ and $h\Delta v_{E_y} = \Delta d_{||}F_{||} - \Delta d_\perp F_\perp$ respectively. Longitudinal fields along the defect's symmetry axis result in equal shifts of all levels, whereas transverse fields split the orbitals into two branches whose energy difference grows with increasing field.[18,19]

We locate two (hh) divacancies located near the center of the 4 lithographically patterned gates approximately 1.5 μm and 3 μm below the surface shown in Fig. 1(b). At T = 10 K, the spin-conserving transitions to the $E_x$ and $E_y$ levels in the excited state are visible in a photoluminescence excitation measurement shown in Fig. 1(c). The biases on each of the metal pads used for applying an electrical field are independently controlled, and the sample is back-mounted to the ground plane by a thin insulating layer of rubber cement. When applying equal bias to all 4 top gates ($V_x = V_y = 0$) relative to the ground plane on the backside of the membrane, an electric field along the symmetry axis (c axis) is applied, shifting both transitions. In Fig. 1(d) and 1(g) we demonstrate DC Stark shifts of the |$A_2$; $S_z$=0> to |$E_x$; $S_z$=0> and |$E_y$; $S_z$=0> transitions respectively of the 1.5 μm deep divacancy by over 2 GHz when applying -300 V to all four top gates. Assuming that the electric field is uniform across the thickness of the sample and taking into account SiC's dielectric constant, we estimate a local field strength of F ≈ 0.25 MV/m at the divacancy and estimate the order of magnitude of $\Delta d_{||}$ to be ≈10 GHz/MV/m, which is similar to the value estimated for the diamond NV center, which was estimated to be within a factor of 3 of $\Delta d_{|| NV}$≈4 GHz/MV/m.[18] In our sample, divacancies in close proximity to each other tend to experience similar values of transverse and longitudinal strain, and divacancies that are closer than 10 microns apart generally have transitions that are separated by <5 GHz (see supplementary material[14]). Therefore, the Stark tuning of a few GHz that we show here is already comparable to the variation in intrinsic strain in our samples, and is sufficient to tune divacancies with different strain so that their photons become indistinguishable.

We also investigate the effect of electric fields transverse to the (hh) divacancy's symmetry axis on the $|A_2; S_z=0>$ to $|E_x; S_z=0>$ and $|E_y; S_z=0>$ transitions in Fig. 1(e) and Fig. 1(h) respectively. We find that the neutral divacancy enters a non-fluorescent state when sufficiently strong (~1 MV/m) transverse electric fields are applied, limiting the range of the Stark shifts generated by transverse electric fields. We note that this is well below SiC's dielectric breakdown field of ~300 MV/m. Quenching of fluorescence of NV centers under large in-plane electric fields has previously been observed, and attributed to electron loss under optical excitation.[20] As shown in Fig. 1(i), the splitting between the two orbital transitions begin to sharply shift near the transition between the bright and dark states. In contrast, for intermediate values of transverse electric fields, the shift is much smaller. The splitting between the two transitions due to intrinsic transverse strain experienced was approximately 30 GHz for both divacancies, though the exact value changed slightly (+/- ~1 GHz) whenever the sample was allowed to warm to room temperature before cooling down again. Similar behavior was observed in the 3 μm deep (hh) divacancy, though the transitions between bright and dark state take place at slightly larger voltages since the lateral electric field is weaker at greater depths.

Even in the absence of transverse electric fields, we observed that when sufficiently positive voltages are equally applied to all four gates relative to the ground plane on the back of the membrane, the divacancies enter a non-fluorescent charge state. We ascribe this dark state to the -1 charge ($VV^{-1}$) state since the population of $VV^{-1}$ is expected to increase at the expense of $VV^0$ with increasingly positive gate voltage. The voltage at which this occurs generally increases with distance to the gate, and we show in Fig 2(a) that charge state of divacancies more than 10 μm away from the gates can be modulated when applying a voltage of +100 V. In contrast, the neutrally charged state persists past $V_z = -300$ V for both divacancies in the center of the device. We ascribe this asymmetry to the slightly n-type nature of the host material, and to the relative location of the $VV^0/VV^{+1}$ and $VV^0/VV^{-1}$ transitions with respect to the Fermi level in our sample.

Since the divacancy ionizes under certain electrical bias conditions, it is important to understand how this constrains the magnitude and direction of the achievable Stark shifts. While the neutral charge state is known to have a spin triplet ground and excited state, enabling it to act as a spin qubit, it is stable only in a limited range of Fermi-level values within the band gap and requires semi-insulating samples. Recent DFT calculations of formation energies suggest that the divacancy in 4H-SiC has four stable charge states within the band gap: +1, 0, −1, and -2.[21] The charge state of the divacancy is influenced by the presence of nearby donor impurities that can also be ionized. Near the Ti-SiC interface, a Schottky barrier is formed whose depth is determined by the applied bias, as soon in Fig. 2(e). Assuming a simple model in which there is uniform ionization of shallow nitrogen donors with density $N_d$, then by solving the Poisson equation the width of the depletion channel is given by $w_d = \sqrt{2\epsilon\epsilon_0|\phi|/eN_d}$ where $\epsilon = 9.6$ is 4H-SiC's dielectric constant, and $\phi$ is the height of the Schottky barrier. In Fig. 2(d) we plot the $VV^0/VV^-$ transition voltage of several defects as a function of their distance from a metal top gate. In gray we show the expected depletion width assuming $N_d = 2.5 \times 10^{14}$ cm$^{-3}$, which is consistent with our measured dopant density. We note that the presence of ionized donors can screen the local electric field at the divacancy, making it difficult to precisely determine the values of the electric dipole moments $\Delta d_{||}$ and $\Delta d_{\perp}$.

The ability to change the charge state of the divacancy can be used to extend nearby nuclear spin coherence times by removing the electronic spin of the divacancy, which acts as a source of decoherence on the nuclei.[22] However, only doubly ionized charge states of the divacancy are expected to be spin-0. While purely optical control has recently been shown to modulate divacancies between neutral and singly charged states[23,24], electrical gating has previously been demonstrated to both singly and doubly ionize diamond NV centers.[22,25,26] The ability to doubly ionize defects is therefore a potential benefit of electrical control over purely optical control of the divacancy charge state.

In order to avoid significant dephasing of nuclear spin memories during the divacancies charge transition, it is important its charge state be changed faster than the hyperfine coupling between the divacancy's electronic spin and the nuclear spin memory. We therefore investigate the limits to this charge transition speed by time-resolved measurements of the fluorescence response to abrupt changes in the applied bias.

The pulse sequence, in which we periodically change between two gate voltages under continuous 975 nm excitation at various optical powers, is shown in Fig. 3(a). At the high bias (8V) the divacancy is in the dark negatively charged state ($VV^{-1}$), whereas at the lower bias (4V) the divacancy is in the bright neutral charge state ($VV^{0}$). After allowing the dynamics to reach a steady state under optical illumination, we change the bias and observe the change in photoluminescence. In Fig. 3(b) and Fig. 3(c) we show the time resolved $VV^{-1}$ to $VV^{0}$ and $VV^{0}$ to $VV^{-1}$ transitions under 2.5 mW $\mu m^{-2}$ 975 nm excitation respectively. We find that the $VV^{0}$ to $VV^{-1}$ transition is well fit by an exponential decay $e^{-t*\gamma_{0-}}$ where $\gamma_{0-}$ is the transition rate from $VV^{0}$ to $VV^{-1}$ and t=0 is the moment the bias switches from 4 V to 8 V. We observe a delay between the $VV^{-1}$ to $VV^{0}$ charge state transition and the bias change, and fit the data to a sigmoid function $(1 + e^{-(t-\tau_{charge})*\gamma_{-0}})^{-1}$ where $\tau_{charge}$ is the delay and $\gamma_{-0}$ is the transition rate from $VV^{-1}$ to $VV^{0}$. Similar delays have been seen in demonstrations of charge state modulation in diamond NV centers[27]. In Fig. 3(e) we measure the transition rates $\gamma_{-0}$ and $\gamma_{0-}$ as a function of 975 nm optical power, finding that the transition rates increase linearly with optical power over the range of powers we used, which was between 0.3 and 15 mW $\mu m^{-2}$. We also investigated the effect of optical power on the delay in Fig. 3(f), finding that it decreases with increasing optical power. In Fig. 3(g), we show that the delay also depends on the bias of the gates, with increasingly positive 'high' gate voltages leading to larger delays for the $VV^{-1}$ to $VV^{0}$ transition when leaving the 'low' gate voltage fixed at 4 V. When using optical powers of 15 mW $\mu m^{-2}$, we were able to modulate both transitions at speeds faster than 5 $\mu s^{-1}$ which is sufficient to prevent significant dephasing during charge state switching for nuclear spin qubits with <1 MHz coupling.

We find that some divacancies in our sample exist in a dark charge state under 975 nm illumination and 0 bias and that these can be brought back into the neutral charge state by applying negative bias to the top gates. We can quantitatively estimate the charge state population of the divacancy as a function of the applied bias using a single shot charge state detection technique in which a low-power laser resonant with one of the $VV^{0}$ GS to ES transitions allows direct and non-destructive readout of the divacancy's charge state.[12,27–29] The fluorescence time trace of one such divacancy at an intermediate voltage near the VV$^0$/VV$^{-1}$ transition is shown in Fig. 4(a), displaying abrupt changes between the bright and dark states. A histogram of the photon count per 8ms readout shows two Poisson distributions, one for the bright state and one for the dark state, whose relative amplitudes can be used to determine the population ratio between VV$^0$ and VV$^{-1}$. When the gate voltage is below the transition, the relative amplitudes of the two Poisson distributions indicate that it is in the neutral charge state close to 100% of the time. This compares favorably to the NV center in diamond, which is left in its spin-1 (NV$^{-1}$ charge) state ~75% of the time after off resonant 532 nm excitation.[12]

We have demonstrated Stark tuning of the neutrally charged divacancy's sharp spin selective optical transitions over several GHz in 4H-SiC. These results indicate that the divacancy's optical resonances can be tuned to compensate for local strain inhomogeneity, and that the range of Stark tuning is already comparable to the variation in resonance frequencies in divacancies in our sample. This provides a practical means to produce identical photons from different divacancies and should find use in tuning divacancies into resonance with optical cavities. We also electrically stabilize the neutral charge state of divacancies several microns away from the top gates even when the divacancies exist naturally in a dark charge state under no applied bias. We used a single shot charge state readout technique to show that the charge state can be prepared with near unity efficiency into the neutral charge state under resonant and off resonant excitation, which compares favorably to the NV center in diamond. These results show that the neutral divacancy, with its electrically tunable spin-selective optical resonances, is a strong candidate for implementing photon mediated quantum information networks using existing fiber optic networks. Furthermore, the fast electrical control of its charge state could enable long lasting nuclear spin memories that are unhindered by magnetic field noise from the neutral divacancy's electronic spin.

This work was supported by the ARL OSD QSEP program, the NSF EFRI 1641099, the AFOSR MURI program, the JSPS KAKENHI(A) 17H01056, the Swedish Research Council (2016-04068), the Carl-Trygger Stiftelse for Vetenskaplig Forskning (CTS 15:339), and the Swedish Energy Agency (43611-1). We thank Alex Crook and Sam Whiteley for assistance in sample preparation.

[1] G. Kucsko, P.C. Maurer, N.Y. Yao, M. Kubo, H.J. Noh, P.K. Lo, H. Park, and M.D. Lukin, Nature **500**, 54 (2013).

[2] G. Balasubramanian, I.Y. Chan, R. Kolesov, M. Al-Hmoud, J. Tisler, C. Shin, C. Kim, A. Wojcik, P.R. Hemmer, A. Krueger, T. Hanke, A. Leitenstorfer, R. Bratschitsch, F. Jelezko, and J. Wrachtrup, Nature **455**, 648 (2008).

[3] F. Dolde, H. Fedder, M.W. Doherty, T. Noebauer, F. Rempp, G. Balasubramanian, T. Wolf, F. Reinhard, L.C.L. Hollenberg, F. Jelezko, J. Wrachtrup, T. Nöbauer, F. Rempp, G. Balasubramanian, T. Wolf, F. Reinhard, L.C.L. Hollenberg, F. Jelezko, and J. Wrachtrup, Nat. Phys. **7**, 459 (2011).

[4] B. Hensen, H. Bernien, A.E. Dréau, A. Reiserer, N. Kalb, M.S. Blok, J. Ruitenberg, R.F.L. Vermeulen, R.N. Schouten, C. Abellán, W. Amaya, V. Pruneri, M.W. Mitchell, M. Markham, D.J. Twitchen, D. Elkouss, S. Wehner, T.H. Taminiau, and R. Hanson, Nature **526**, 682 (2015).

[5] P.G. Baranov, A.P. Bundakova, A.A. Soltamova, S.B. Orlinskii, I. V Borovykh, R. Zondervan, R. Verberk, and J. Schmidt, Phys. Rev. B **83**, 125203 (2011).

[6] W.F. Koehl, B.B. Buckley, F.J. Heremans, G. Calusine, and D.D. Awschalom, Nature **479**, 84 (2011).

[7] B. Pingault, D. Jarausch, C. Hepp, L. Klintberg, and J.N. Becker, Nat. Commun. **8**, 15579 (2017).

[8] D.J. Christle, P. V Klimov, C.F. de las Casas, K. Szász, V. Ivády, V. Jokubavicius, J.U. Hassan, M. Syväjärvi, W.F. Koehl, T. Ohshima, N.T. Son, E. Janzén, Á. Gali, and D.D. Awschalom, Phys. Rev. X **7**, 21046 (2017).

[9] H. Bernien, B. Hensen, W. Pfaff, G. Koolstra, M.S. Blok, L. Robledo, T.H. Taminiau, M. Markham, D.J. Twitchen, L. Childress, and R. Hanson, Nature **497**, 86 (2013).

[10] A. Faraon, C. Santori, Z. Huang, V.M. Acosta, and R.G. Beausoleil, Phys. Rev. Lett. **109**, 33604 (2012).

[11] G. Calusine, A. Politi, and D.D. Awschalom, Appl. Phys. Lett. **105**, 1 (2014).

[12] N. Aslam, G. Waldherr, P. Neumann, F. Jelezko, and J. Wrachtrup, New J. Phys. **15**, 13064 (2013).

[13] J. Hassan, J.P. Bergman, A. Henry, and E. Janzén, J. Cryst. Growth **310**, 4424 (2008).

[14] See supplementary material for details.

[15] J.R. Maze, A. Gali, E. Togan, Y. Chu, A. Trifonov, E. Kaxiras, and M.D. Lukin, New J. Phys. **13**, 25025 (2011).

[16] M.W. Doherty, N.B. Manson, P. Delaney, and L.C.L. Hollenberg, New J. Phys. **13**, 25019 (2011).

[17] A.L. Falk, P. V Klimov, B.B. Buckley, V. Ivady, I.A. Abrikosov, G. Calusine, W.F. Koehl, Á. Gali, and D.D. Awschalom, Phys. Rev. Lett. **112**, 187601 (2014).

[18] V.M. Acosta, C. Santori, A. Faraon, Z. Huang, K.C. Fu, A. Stacey, D.A. Simpson, K. Ganesan, A.D. Greentree, S. Prawer, and R.G. Beausoleil, Phys. Rev. Lett. **108**, 206401 (2012).

[19] L.C. Bassett, F.J. Heremans, C.G. Yale, B.B. Buckley, and D.D. Awschalom, Phys. Rev. Lett. **107**, 266403 (2011).

[20] M. V Hauf, P. Simon, N. Aslam, M. Pfender, P. Neumann, S. Pezzagna, J. Meijer, J. Wrachtrup, M. Stutzmann, F. Reinhard, and J.A. Garrido, Nano Lett. **14**, 2359 (2014).

[21] L. Gordon, A. Janotti, and C.G. Van De Walle, Phys. Rev. B **92**, 45208 (2015).

[22] M. Pfender, N. Aslam, P. Simon, D. Antonov, G.O. Thiering, S. Burk, F. Fávaro De Oliveira, A. Denisenko, H. Fedder, J. Meijer, J.A. Garrido, A. Gali, T. Teraji, J. Isoya, M.W. Doherty, A. Alkauskas, A. Gallo, A. Grüneis, P. Neumann, and J. Wrachtrup, arXiv:1702.01590 (2017).

[23] D.A. Golter and C.W. Lai, arXiv:1707.01558 (2017).

[24] G. Wolfowicz, C.P. Anderson, A.L. Yeats, S. Whiteley, J. Niklas, O.G. Poluektov, F.J. Heremans, and D.D. Awschalom, arXiv:1705.09714 (2017).

[25] B. Grotz, M. V Hauf, M. Dankerl, B. Naydenov, S. Pezzagna, J. Meijer, F. Jelezko, J. Wrachtrup, M. Stutzmann, F. Reinhard, and J.A. Garrido, Nat. Commun. **3**, 729 (2012).

[26] C. Schreyvogel, V. Polyakov, R. Wunderlich, J. Meijer, and C.E. Nebel, Sci. Rep. **5**, 12160 (2015).

[27] Y. Doi, T. Makino, H. Kato, D. Takeuchi, M. Ogura, H. Okushi, H. Morishita, T. Tashima, S. Miwa, S.

Yamasaki, P. Neumann, J. Wrachtrup, Y. Suzuki, and N. Mizuochi, Phys. Rev. X **4**, 11057 (2014).

[28] G. Waldherr, P. Neumann, S.F. Huelga, F. Jelezko, and J. Wrachtrup, Phys. Rev. Lett. **107**, 90401 (2011).

[29] Y. Doi, T. Fukui, H. Kato, T. Makino, S. Yamasaki, T. Tashima, H. Morishita, S. Miwa, F. Jelezko, Y. Suzuki, and N. Mizuochi, Phys. Rev. B **93**, 81203 (2016).

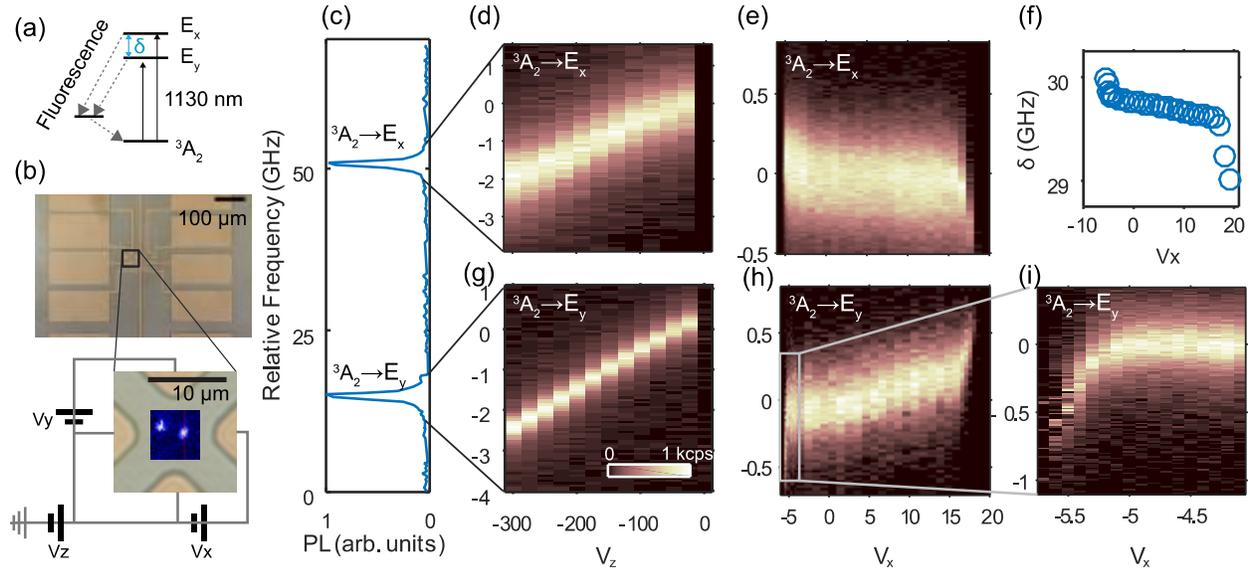

FIG. 1. (a) Energy-level diagram showing the orbital levels that are excited by resonant excitation (black arrows). (b) Device for applying lateral and vertical electric fields with 10/100 nm Ti/Au electrodes on 120 μm thick membrane of 4H-SiC. Photoluminescence of two divacancies measured in the center of this device are shown in the inset. (c) Photoluminescence excitation spectra of a single divacancy showing strain splitting between the $^3A_2 \rightarrow E_x$ and $^3A_2 \rightarrow E_y$ transitions taken at 10 K. A narrow line resonant laser is swept over the transitions while photons in the red shifted phonon sideband are collected. The frequency is relative to 264830 GHz. (d) DC Stark tuning of the $^3A_2 \rightarrow E_x$ and (g) $^3A_2 \rightarrow E_y$ transition with application of $V_z$ ($V_x = V_y = 0$) relative to the ground plane on the membrane's backside. Electric fields applied along the symmetry axis of the divacancy shift both branches in the same direction. Applying positive voltages results in the charge state of the divacancy changing. (e) PLE spectra of the $^3A_2 \rightarrow E_x$ and (h) $^3A_2 \rightarrow E_y$ transition as a function of lateral bias $V_x$ with $V_z = -20V$. (i) Magnified region in (h) showing the transition between the bright and dark states. (f) The splitting between the $^3A_2 \rightarrow E_x$ and $^3A_2 \rightarrow E_y$ transitions as a function of transverse voltage.

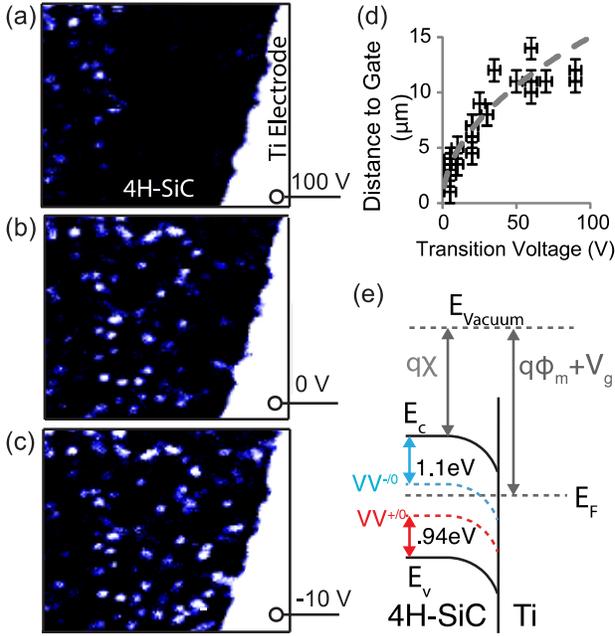

FIG 2. (a) Photoluminescence map of divacancies near the SiC-Ti interface at (a) 100V, (b) 0V, and (c) -10V showing divacancies entering a non-fluorescent charge state. (d) Distance dependence of the $VV^0/VV^-$ transition voltage of several divacancies. The gray dashed line corresponds to the expected depletion width assuming ionization of dopants with density of 2.5 x $10^{14}$ cm$^{-3}$. (e) Band diagram of the SiC-Ti interface illustrating band bending where $\chi$ is the electron affinity of 4H-SiC (4.17 eV), $\phi_m$ is the work function (4.3 eV for Ti), $V_g$ is the applied bias on the gate, $E_v$ is the valence band maximum, $E_C$ is the conduction band minimum, and $E_F$ is the Fermi level. Also shown are the charge transition between $VV^{+/0}$ and $VV^{-/0}$ calculated in Ref. 21.

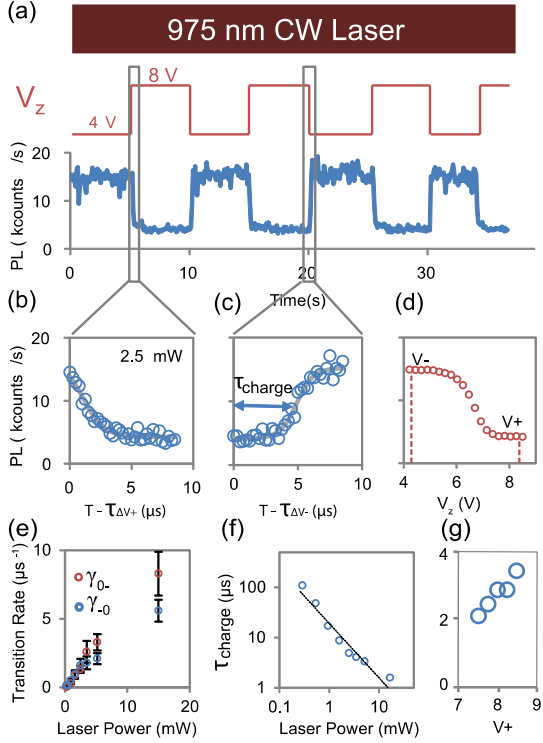

FIG 3. (a) Charge state modulation of a single divacancy by applying bias to an electrode 3 μm away that forms a Schottky barrier with the 4H-SiC. (b) Dynamics of the $VV^0$ to $VV^-$ state transition. The data is fit to an exponential decay $e^{-\gamma_{0-}t}$ where $\gamma_{0-}$ is the transition rate from $VV^0$ to $VV^-$. (c) Dynamics of the $VV^-$ to $VV^0$ transition. The data is fit to a sigmoid function $(1 + e^{-(t-\tau_{charge})*\gamma_{-0}})^{-1}$ where $\tau_{charge}$ is the delay between the abrupt voltage change and when the charge transition occurs, and $\gamma_{-0}$ is the transition rate from $VV^{-1}$ to $VV^0$. (d) Photoluminescence of a divacancy as a function of voltage showing $VV^{-1}$ to $VV^0$ transition of +6 V. (e) Transition rates determined from fits to exponential decay and sigmoid function for the $VV^0$ to $VV^-$ transition (red circles) and $VV^{-1}$ to $VV^0$ (blue circles) respectively as a function of laser power. Error bars are the 95% confidence interval. (f) Charging time of the $VV^{-1}$ to $VV^0$ transition decreases with increasing laser power. (g) We also observe that the charging time also increases as V+ is increased for a fixed laser power of 5.2 mW while V- is held fixed at 4 V.

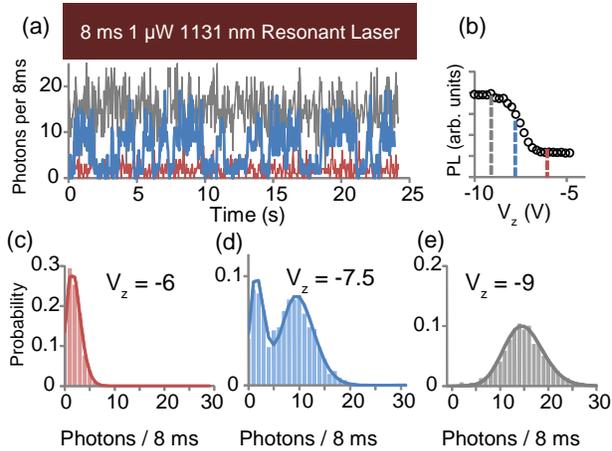

FIG 4. (a) Time trace of the fluorescence of a single divacancy under continuous 1μW resonant excitation when the gate is biased -9V (grey), -7.5V (blue), and -6V (red). The divacancy is initially in a dark charge state unless a sufficiently negative bias is applied to the gate. At -7.5 V, the divacancy can be observed to abruptly jump between a bright ($VV^0$) and dark ($VV^-$) state. (b) Photoluminescence as a function of gate voltage under continuous 1 mW 975 nm (off resonant) excitation of a divacancy that is in a dark charge state at 0 bias and only enters the neutral charge state when a sufficiently negative bias is applied to a surface electrode 3 μm away. Histogram of number of photons per 8 ms during 1 μW resonant excitation at (c) -7.5 V, (d) -6 V, (e) -9 V showing the divacancy is stabilized into the neutral state with application of sufficiently negative top gate voltage.